# Exploring differences in injury severity between occupant groups involved in fatal rear-end crashes: A correlated random parameter logit model with mean heterogeneity

**Renteng Yuan (First author)**
Jiangsu Key Laboratory of Urban ITS
School of Transportation
Southeast University, Nanjing, Jiangsu, P. R. China, and 210000
Email: rtengyuan123@126.com
**Xin Gu**
Beijing Key Laboratory of Traffic Engineering
Beijing University of Technology, Beijing, P. R. China, and 100124
Email: guxin@bjut.edu.cn
**Zhipeng Peng**
College of Transportation Engineering
Chang'an University, Xi'an, Shaanxi, P. R. China, and 710000
Email: pengzhipengcdlw@163.com
**Qiaojun Xiang (Corresponding author)**
Jiangsu Key Laboratory of Urban ITS
School of Transportation
Southeast University, Nanjing, Jiangsu, P. R. China, and 210000
Email: xqj@seu.edu.cn

**Abstract**
Rear-end crashes are one of the most common crash types. Passenger cars involved in rear-end crashes frequently produce severe outcomes. However, no study investigated the differences in the injury severity of occupant groups when cars are involved as following and leading vehicles in rear-end crashes. Therefore, the focus of this investigation is to compare the key factors affecting the injury severity between the font- and rear-car occupant groups in rear-end crashes. First, data is extracted from the Fatality Analysis Reporting System (FARS) for two types of rear-end crashes from 2017 to 2019, including passenger cars as rear-end and rear-ended vehicles. Significant injury severity difference between font- and rear-car occupant groups is found by conducting likelihood ratio test. Moreover, the font- and rear-car occupant groups are modelled by the correlated random parameter logit model with heterogeneity in means (CRPLHM) and the random parameter logit model with heterogeneity in means (RPLHM), respectively. Overall, significant differences in the factors affecting the injury severity of different occupant groups are found in rear-end crashes, such as occupant positions, driver age, overturn, vehicle type, etc. For instance, the driving and front-right positions significantly increase the probability of severe injury when struck by another vehicle. Large truck-strike-car tends to cause severe outcomes compared to car-strike-large truck. This study provides an insightful knowledge of mechanism of occupant injury severity in rear-end crashes, and could help traffic managers implement effective countermeasures to mitigate the crash severity.

## 1. Introduction

Passenger cars have become the primary mode of transportation for people because they are flexible, fast, and comfortable(LindaSTEG; Stradling et al., 2007). According to the Federal Highway Administration (FHWA), passenger cars accounted for more than half of annual vehicle distance traveled in the United States. Although passenger cars bring many conveniences to life, the hazards caused by them also cannot be ignored. Previous studies have indicated that passenger cars are more likely to experience severe crashes when involved in crashes (Abdel-Aty, 2003; Eluru et al., 2010; Ulfarsson & Mannering, 2004; Yu, R. & Abdel-Aty, 2014). According to NHTSA (https://cdan.dot.gov/tsftables/tsfar.htm#), rear-end crashes accounted for 71.4% of multiple-vehicle fatal crashes involving passenger vehicles in 2019(NHTSA). Therefore, in-depth analysis of the risk factors of occupant (driver and passenger) injuries in rear-end crashes is an important prerequisite for making effective safety improvement strategies.

A rear-end crash commonly occurs when a following (rear) vehicle collides with a leading (front) vehicle. The vehicles involved in rear-end crashes evince obvious differences such as vehicle type, braking performance, driving speed, etc. Previous study has proved that the outcome of car-strike-truck crashes and truck-strike-car crashes was quite different in rear-end crashes(Shao et al., 2020). Occupants can be classified into two groups based on the role of the vehicle in rear-end crashes (the front vehicle group and the rear vehicle group)(Yuan, Q. et al., 2017). Meanwhile, significant differences in the factors affecting the injury severity of rear- and front-drivers involved have been reported in the same rear-end crash (Chen, F. et al., 2019; Dabbour et al., 2020). Multiple occupants (driver and passenger) may be involved in a rear-end crash. Hence, it can be inferred that if such heterogeneity between occupant groups is ignored in model development and analysis, it may lead to biased estimation. The contributing mechanism and heterogeneous effects of factors on the injury severity of different occupant groups in rear-end crashes have not been fully explored.

On the other hand, the drivers' injury severity has been widely treated as a dependent variable in previous studies (Champahom et al., 2020; Chen, C., Zhang, Liu, et al., 2016; Chen, C. et al., 2015; Chen, F. et al., 2019; Dabbour et al., 2020). Only limited studies have been conducted to investigate the factors affecting injury severity at occupant-level (Haq et al., 2020; Kidando, 2022; Meng et al., 2017; Zhu & Srinivasan, 2011). If the analysis is limited on driver level, not only would this be a waste of a limited sample, but it would also be difficult to explain why different occupants suffered different levels of injury in the same rear-end crash. This may further affect the effectiveness of management's traffic safety improvement strategies.

To this end, the study aims to employ the correlated mixed logit models with heterogeneity in means (CRPLHM) to investigate the significant factors determining the injury severity of front-and rear-car occupant groups in rear-end crashes. To our knowledge, no study has separately investigated separately the differences between front- and rear-car occupant injury severity in rear-end crashes. The rest of the paper is organized as follows. First, a literature review is presented in Section 2. Second, the employed dataset is introduced in Section 3. Subsequently, Section 4 introduces the detailed methodology design. Next, research results are shown in Section 5. Finally, the discussions and conclusions are summarized in Section 6.

## 2. Literature review

Rear-end crash analysis has attracted worldwide attention(Ahmadi et al., 2018; Mohamed Shawky A). Previous studies mainly focused on the following aspects for analyzing the severity of rear-end crashes: the first aspect is to investigate rear-end crashes occurred on different road segments. For instance, Wang and Luo(Wang Yonggang 2021)analyzed the severity of rear-end crashes occurred on a mountainous expressway. Yu et al. (Yu, M., Ma, et al., 2020)and Champahom et al.(Champahom et al., 2020) compared various factors affecting the likelihood of rear-end crash severity in urban and rural areas. Yu et al.(Yu, M., Zheng, et al., 2020), Zhang and Hassan (Zhang & Hassan, 2019) focused on investigating the factors contributing to injury severity in work zone rear-end crashes. The second aspect focuses on rear-end crashes with special-vehicle-involved, especially those truck-involved, to investigate the contributing factors associated with rear-end crash severity. For example, Liu and fan (Liu & Fan, 2020) investigate the critical factors affecting the severity of large truck related rear-end crashes. Peng et al. (Peng, Z. et al., 2021) developed ordered logistic regression models, to reveal the differences in the injury severity of drivers and copilots in rear-end crashes on expressways, and the results showed that the copilots were more vulnerable in truck rear-end crashes compared with drivers. Yuan et al. (Yuan, Q. et al., 2017) investigated occupant injury severity in rear-end crashes involving trucks as the front vehicle, and found that trucks as the front vehicles tend to significantly increase the likelihood of fatal injury.

Crash severity is influenced by driver characteristics, vehicle characteristics, road characteristics, and environmental characteristics. In previous studies, drivers or vehicle characteristics were generally considered as one variable to explore their effect on rear-end crashes injury severity (take "if large trucks are involved in rear-end crashes," or "if older drivers are involved in rear-end crashes," as a filter criteria in crash data) (Peng, Z. et al., 2021; Yuan, R. et al., 2022; Yuan, Y. et al., 2021). These aggregation treatments might ignore the heterogeneity among variables and lead to some controversial conclusions. For instance, some studies believe that the involvement of older drivers in rear-end crashes may decrease the probability of fatal injury due to their more experienced driving skill(Zhang & Hassan, 2019) (Yuan, Q. et al., 2017), while some other studies found that older driver may increase the risk of death in the rear-end collisions due to their long response time (Ahmadi et al., 2018; Yu, M., Zheng, et al., 2020). Similar mixed conclusions were also found regarding the effect of large trucks on the severity of rear-end collisions. For example, Liu and fan(Liu & Fan, 2020) and Chen et al. (Chen, C. et al., 2015) concluded that large trucks may increase collision severity, while Chen et al.(Chen, C., Zhang, Huang, et al., 2016) and Huang et al. (Huang et al., 2008) found that large trucks have better resistance to severe crashes, and therefore, collision severity is decreased. Hence, it is necessary to refine the variable groupings for in-depth analysis, particularly in rear-end crashes, which may have heterogeneous effects on occupant injury severity.

From a methodological viewpoint, multinomial logit (MNL) models have been widely used to estimate the risk factors of crash severity. However, they are limited by the independence of irrelevant alternatives (IIA) assumption (Dabbour et al., 2020). Indeed, risk factors involved in crashes may be correlated, and obtaining all the data that may determine injury severity is difficult. The absence of such the correlations and important data can potentially lead to biased and inconsistent parameter estimates(Mannering et al., 2016). To mitigate this effect, random parameter logit (RPL) models were further developed to capture unobserved heterogeneity, which allow the partial parameters vary across each observation (Islam & Burton, 2019; Liu & Fan, 2020; Uddin & Huynh, 2020). Moreover, random parameter logit models with mean-variance heterogeneity were employed to explain the heterogeneity in means and variances of random parameters, which provide a superior goodness-of-fit and offer more insights into factors of crash severity(Hou et al., 2019; Mannering et al., 2016). In order to analyze the association between random parameters, a correlated random parameters model has been conducted by many studies(Saeed et al., 2019; Se, Chamroeun et al., 2021; Wang, K. et al., 2021). In correlated random parameters approach, the random parameters are allowed to be correlated (as opposed to the uncorrelated structure in conventional random parameters model). This modeling approach extends the frontier of the conventional random parameter logit models by relaxing the likely correlations among the random parameters.

## 3 Data description

This study focuses on two-vehicle rear-end crashes involving at least one passenger car. Using crash data is extracted from the United States between 2017 and 2019 available from the Fatality Analysis Reporting System (FARS) (ftp://ftp.nhtsa.dot.gov/fars/). Each crash involves at least 1 occupant death within 30 days. The FARS files are obtained from various states' documents, such as: police crash reports, death certificates, coroner/medical examiner reports, etc. The dataset contains information regarding crash, environment, vehicle, passenger and driver characteristics, regarding crash cause and time, road surface material, weather, vehicle type and crash point, occupant position, driver age and sex information.

Furthermore, vehicle and driver characteristics are integrated into front vehicle and driver characteristics, rear vehicle and driver characteristics based on crash point. Those samples lacking front-and rear-vehicle and driver characteristics are excluded. And samples involving variables used that are recorded as unknown or not reported are also removed. Ultimately, a total of 1343 crash records contained information on 2080 occupants (both driver and passenger) are selected. Since the study aims to investigate the difference in factors determining the injury severity of front-and rear-car occupant groups, the dataset is divided into the front-car occupant group (FV dataset) and the rear-car occupant group (RV dataset). For RV dataset, the rear vehicle is limited to a passenger car, while the front vehicle is perhaps a large truck, a SUV, or a passenger car, etc. Similarly, in FV dataset, the front vehicle is limited to a passenger car, while the rear vehicle type is unrestricted. The RV dataset contains 596 crash records and 1002 occupants. The FV dataset contains 747 crash records and 1078 occupants.

The front- and rear-car occupant injury severities are treated as the dependent variables in the FV and RV datasets, respectively. To obtain a sufficient sample size in each level of the response variables(Wang, K. et al., 2019; Zeng et al., 2022), the crash data is aggregated into three injury severity categories: minor injury (original category: no apparent injury and possible injury), serious injury (original category: suspected minor and suspected serious injury), and fatal injury (original categories: death). Samples reported as injured but of unknown severity were removed. **Table1** summarizes the descriptive statistics of crash data. The spatial distribution of extracted rear-end crashes is shown in **Fig 1**.

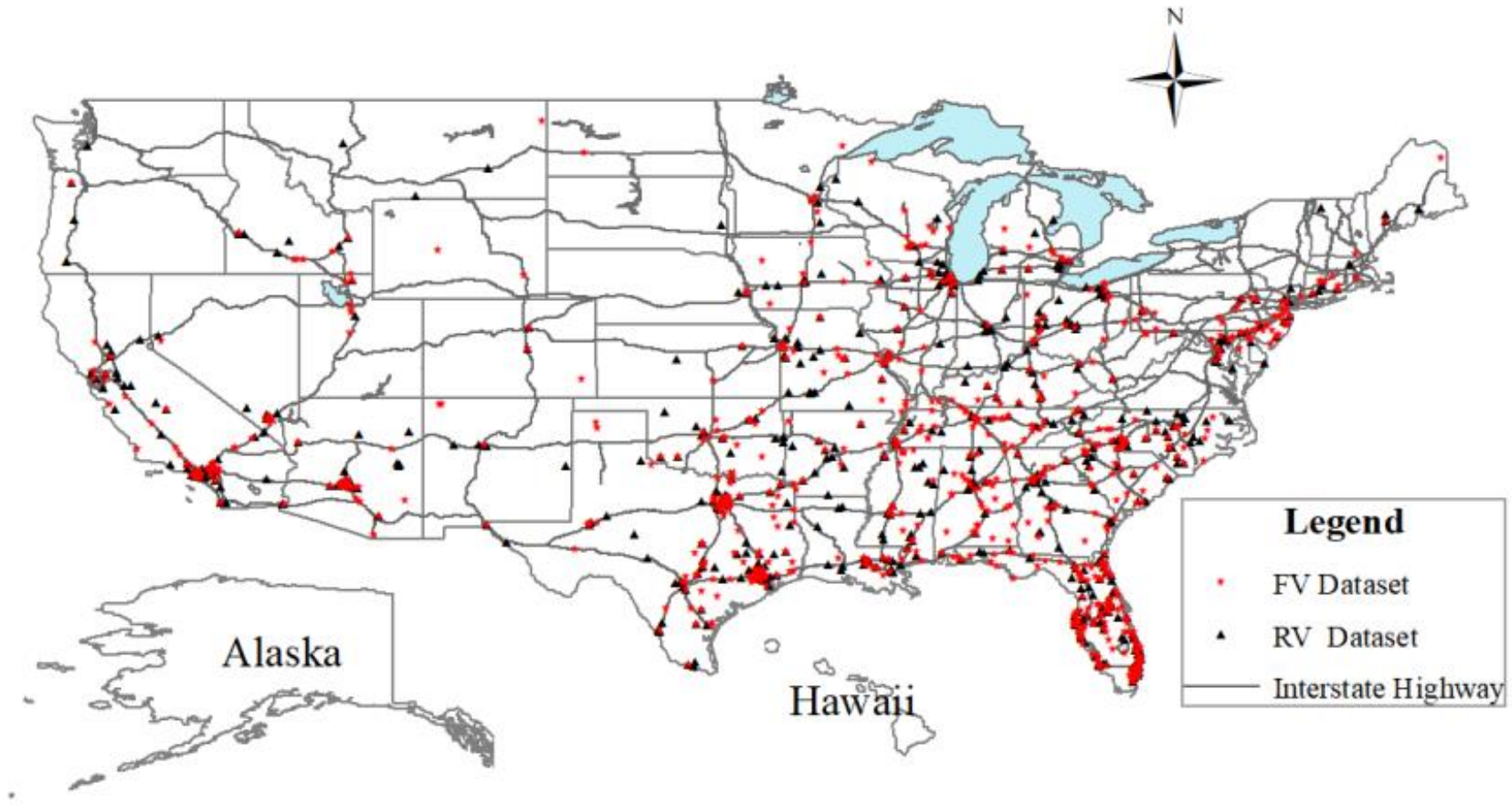

**Fig 1 Spatial distribution of extracted rear-end crashes**

Note. Hawaii (HI) and Alaska (AK) were shifted and not to scale.

## Table 1 Statistical-descriptive analysis of rear-end crashes

| Variable | | Description | FV Dataset (N = 1078) | | | | | | RV Dataset (N = 1002) | | | | | |
|---|---|---|---|---|---|---|---|---|---|---|---|---|---|---|
| | | | Fatal injury | | Serious injury | | Minor injury | | Fatal injury | | Serious injury | | Minor injury | |
| | | | n | % | n | % | n | % | N | % | n | % | n | % |
| **Occupant and crash characteristics** | | | | | | | | | | | | | | |
| Safety belt | Safety belt | 1 if the occupant used the safety belt, 0 otherwise. | 337 | 38.9 | 250 | 28.8 | 280 | 32.3 | 266 | 41.9 | 178 | 28.0 | 191 | 30.1 |
| Seat position | Driver position | 1 if the occupant at driver position, 0 otherwise. | 291 | 49.8 | 134 | 22.9 | 159 | 27.2 | 426 | 57.3 | 159 | 21.4 | 159 | 21.4 |
| | Front-right | 1 if the occupant at first row right position, 0 otherwise | 81 | 31.2 | 110 | 42.3 | 69 | 26.5 | 70 | 46.1 | 51 | 33.6 | 31 | 20.4 |
| | Second * | 1 if the occupant at second row, 0 otherwise. | 104 | 44.4 | 60 | 25.6 | 70 | 29.9 | 27 | 23.3 | 49 | 42.2 | 40 | 34.5 |
| Collision Cause | Stopped | 1 if the rear-end crashes occurred due to the front vehicles stopping, 0 otherwise. | 192 | 48.9 | 113 | 28.8 | 88 | 22.4 | 188 | 60.6 | 77 | 24.8 | 45 | 14.5 |
| | Slower | 1 if the rear-end crashes occurred due to the slower speed of the front vehicles, 0 otherwise. | 160 | 36.9 | 115 | 26.5 | 159 | 36.6 | 226 | 46.7 | 132 | 27.3 | 126 | 26.0 |
| | Decelerating | 1 if the rear-end crashes occurred due to the front vehicles decelerating, 0 otherwise. | 64 | 45.1 | 50 | 35.2 | 28 | 19.7 | 69 | 59.0 | 27 | 23.1 | 21 | 17.9 |
| | Other * | 1 if the rear-end crashes occurred due to the vehicle out of control, changing lanes, accelerating, etc. 0 otherwise. | 60 | 55.0 | 26 | 23.9 | 23 | 21.1 | 40 | 44.0 | 23 | 25.3 | 28 | 30.8 |
| **Environmental characteristics** | | | | | | | | | | | | | | |
| Light condition | Lighted | 1 if the rear-end crashes occurred at night with lights, 0 otherwise. | 64 | 31.2 | 46 | 22.4 | 95 | 46.3 | 127 | 55.2 | 66 | 28.7 | 37 | 16.1 |
| | Dark | 1 if the rear-end crashes occurred at night without lights,0 otherwise. | 160 | 46.6 | 97 | 28.3 | 86 | 25.1 | 154 | 47.1 | 92 | 28.1 | 81 | 24.8 |
| | Daylight * | 1 if the rear-end crashes occurred at daylight, 0 otherwise. | 221 | 41.7 | 161 | 30.4 | 148 | 27.9 | 242 | 54.4 | 101 | 22.7 | 102 | 22.9 |
| Surface type | Concrete | 1 if crash occurred at concrete road, 0 otherwise. | 52 | 43.7 | 27 | 22.7 | 40 | 33.6 | 51 | 53.7 | 26 | 27.4 | 18 | 18.9 |
| | Asphalt | 1 if crash occurred at asphalt road, 0 otherwise. | 264 | 42.9 | 168 | 27.3 | 184 | 29.9 | 273 | 52.2 | 133 | 25.4 | 117 | 22.4 |
| | Others * | 1 if crash occurred at slag, gravel or stone road, 0 otherwise. | 160 | 46.6 | 109 | 31.8 | 74 | 21.6 | 199 | 51.8 | 100 | 26.0 | 85 | 22.1 |
| Road condition | | 1 if wet, snowy or icy surface conditions, 0 otherwise. | 69 | 47.6 | 34 | 23.4 | 42 | 29.0 | 48 | 50.0 | 23 | 24.0 | 25 | 26.0 |
| Weather | Adverse weather | 1 if weather condition was rain, snow, fog, 0 otherwise. | 131 | 46.2 | 78 | 27.5 | 74 | 26.3 | 114 | 47.6 | 66 | 27.6 | 59 | 24.8 |
| Weekend | Weekend | 1 if the rear-end crash occurred on a weekend , 0 weekday | 213 | 42.2 | 141 | 27.9 | 150 | 29.9 | 172 | 46.1 | 107 | 28.6 | 94 | 15.3 |
| **Front vehicle and driver characteristics** | | | | | | | | | | | | | | |
| | Young driver | 1if the front vehicle driver is under 25 years,0 otherwise | 101 | 41.9 | 75 | 31.1 | 65 | 27.0 | 46 | 38.3 | 40 | 33.3 | 34 | 28.3 |
| Driver age | Middle driver* | 1if the front vehicle driver is between 25 - 60 years, 0 otherwise | 241 | 41.3 | 158 | 27.1 | 185 | 31.7 | 398 | 58.5 | 160 | 23.5 | 122 | 17.9 |
| | Older driver | 1if the front vehicel driver is above 60 years, 0 otherwise | 134 | 53.0 | 71 | 28.1 | 48 | 19.0 | 79 | 39.1 | 59 | 29.2 | 64 | 31.7 |
| Sex | Male driver | 1if the front vehicle driver is male, 0 female | 284 | 46.8 | 158 | 26.0 | 165 | 27.2 | 445 | 56.7 | 186 | 23.7 | 154 | 19.6 |
| | Passenger car* | 1if the front vehicle is a passenger car, 0 otherwise | 476 | 44.2 | 304 | 28.2 | 298 | 27.6 | 80 | 26.4 | 111 | 36.6 | 112 | 37.0 |
| | Pickup | 1if the front vehicle is a pickup, 0 otherwise | -- | -- | -- | -- | -- | -- | 50 | 39.7 | 35 | 27.8 | 41 | 32.5 |
| Vehicle type | Van | 1if the front vehicle is a van, 0 otherwise | -- | -- | -- | -- | -- | -- | 18 | 41.9 | 9 | 20.9 | 16 | 37.2 |
| | SUV | 1if the front vehicle is a suv, 0 otherwise | -- | -- | -- | -- | -- | -- | 59 | 40.7 | 41 | 28.3 | 45 | 31.0 |
| | Large Truck | 1if the front vehicle is a large truck, 0 otherwise | -- | -- | -- | -- | -- | -- | 334 | 79.1 | 66 | 15.6 | 22 | 5.2 |
| Overturn | Overturn | 1if the front vehicle overturned, 0 otherwise | 44 | 47.8 | 36 | 39.1 | 12 | 13.0 | 13 | 9.6 | 46 | 34.1 | 76 | 56.3 |
| Disable damage | Disable damage | 1if the front vehicle is disable damage, 0 otherwise | 459 | 47.5 | 297 | 30.7 | 210 | 21.7 | 221 | 37.4 | 184 | 31.1 | 186 | 31.5 |
| **Rear vehicle and driver characteristics** | | | | | | | | | | | | | | |
| | Young driver | 1 if the rear vehicle driver age is under 25 years, otherwise | 118 | 50.4 | 75 | 32.1 | 41 | 17.5 | 139 | 47.6 | 82 | 28.1 | 71 | 24.3 |
| Driver age | Middle driver * | 1 if the rear vehicle driver age is between 25 - 60 years, otherwise | 305 | 46.7 | 172 | 26.3 | 176 | 27.0 | 285 | 49.8 | 153 | 26.7 | 134 | 23.4 |
| | Older driver | 1 if the rear vehicle driver age is above 60 years1, otherwise | 53 | 27.7 | 57 | 29.8 | 81 | 42.4 | 99 | 71.7 | 24 | 17.4 | 15 | 10.9 |
| Sex | Male driver | 1if the rear vehicle driver is male, 0 female | 379 | 45.6 | 244 | 29.3 | 209 | 25.1 | 398 | 55.7 | 175 | 24.5 | 141 | 19.7 |
| Distracted driving | Distracted driving | 1if the rear driver is distracted driving, 0 otherwise | 99 | 45.4 | 69 | 31.7 | 50 | 22.9 | 95 | 49.0 | 50 | 25.8 | 49 | 25.3 |
| | Passenger car* | 1if the rear vehicle is a passenger car, 0 otherwise | 159 | 37.8 | 108 | 25.7 | 154 | 36.6 | 523 | 52.2 | 259 | 25.8 | 220 | 22.0 |
| | Pickup | 1if the rear vehicle is a pickup, 0 otherwise | 112 | 49.1 | 67 | 29.4 | 49 | 21.5 | -- | -- | -- | -- | -- | -- |
| Vehicle type | Van | 1if the rear vehicle is a van, 0 otherwise | 19 | 40.4 | 18 | 38.3 | 10 | 21.3 | -- | -- | -- | -- | -- | -- |
| | SUV | 1if the rear vehicle is a suv, 0 otherwise | 93 | 43.7 | 62 | 29.1 | 58 | 27.2 | -- | -- | -- | -- | -- | -- |
| | Large Truck | 1if the rear vehicle is a large tuck, 0 otherwise | 93 | 55.0 | 49 | 29.0 | 27 | 16.0 | -- | -- | -- | -- | -- | -- |
| Overturn | Overturn | 1if the rear vehicle overturned, 0 otherwise | 41 | 27.9 | 41 | 27.9 | 65 | 44.2 | 54 | 52.9 | 41 | 40.2 | 7 | 6.9 |
| Disable damage | Disable damage | 1if the rear vehicle is disable damage, 0 otherwise | 404 | 44.2 | 254 | 27.8 | 255 | 27.9 | 509 | 53.2 | 252 | 26.4 | 195 | 20.4 |

Note: variables marked with “*” are base variables in the model.

# 4 Methodology

4.1 Correlated random parameters model with heterogeneity in mean (CRPLHM)

To capture unobserved heterogeneity, this paper adopts correlated random parameter models which consider the correlation among random parameters and possible heterogeneity in means. The general modelling framework is defined starting with Multinomial Logit model. Initially, a linear function for each observation *n* is defined as follows(Washington, 2020):

$$Y_{in} = \beta_i X_{in} + \varepsilon_{in} \tag{1}$$

Where $Y_{in}$ is the function of crash characteristics which determines the propensity of attaining injury severity category i in crash n, $X_{in}$ is the vector of explanatory variables for the $n^{th}$ crash with the $i^{th}$ outcome, $\beta_i$ is a vector of estimable parameters for the $i^{th}$ explanatory variable, and $\varepsilon_{in}$ is the error term assumed to be the generalized extreme value distributed. The Multinomial Logit model can be composed as:

$$P_n(i) = \frac{EXP[\beta_i X_{in}]}{\sum_{i=1}^{I} EXP[\beta_i X_{in}]} \tag{2}$$

Where $P_n(i)$ represents the probability of injury severity category i in crash n, *I* represents the number of samples. Due to the unobserved heterogeneity, it should note that some $\beta_i$ can be fixed across the samples and some vary over samples randomly. A mixed logit model was developed, which allows some coefficients follow normal, lognormal, Weibull, and uniform distributions. The probabilities in mixed logit can be expressed as:

$$P_n(i|\varphi) = \int \frac{\exp[\beta_i X_{in}]}{\sum_{i=1}^{I} \exp[\beta_i X_{in}]} f(\beta_i|\varphi) d_{\beta_i} \tag{3}$$

Where $f(\beta_i|\varphi)$ is the density function of $\beta_i$, $\varphi$ is a vector of parameters (mean and variance) that describe the density function. In this paper random parameters are specified to follow a normal distribution. To further explain the unobserved heterogeneity, $\beta_i$ can be accounted for as follows:

$$\beta_i = \beta + \delta Z_i + \Gamma v_i \tag{4}$$

Where $\beta$ denotes the mean value of the random parameter vector, $Z_i$ represents a vector of explanatory determining the means of the random parameters, i.e., the variables capturing the heterogeneity in the mean with a vector of coefficients $\delta$, $v_i$ represents a random term which follow a normal distribution, $\Gamma$ is a symmetric Cholesky matrix, that was employed to compute the standard deviation of the random parameters (Saeed et al., 2019; Se, Chamroeun et al., 2021). The covariance matrix-V of random parameters can be expressed as:

$$V = \Gamma\Gamma' \tag{5}$$

The traditional RPL model specified thus far assumes that the random parameters are uncorrelated. $\Gamma$ is a diagonal matrix, and the variance of $\beta_i$ is simply $\sigma_i^2$. However, previous study has showed that the heterogeneity captured by the random parameters may be correlated rather than independent(Fountas et al., 2021). To allow free correlation among the random parameters, a more generalized formulation of the Cholesky matrix, Γ, is adopted, with the off-diagonal elements taking non-zero values, and was shown in Eq. (6).

$$\Gamma=\begin{bmatrix} \sigma_{1,1} & \sigma_{1,2} & \cdots & \sigma_{1,m-1} & \sigma_{1,m} \\ \sigma_{2,1} & \sigma_{2,2} & \cdots & \sigma_{2,m-1} & \sigma_{2,m} \\ \vdots & \vdots & \ddots & \vdots & \vdots \\ \sigma_{m-1,1} & \sigma_{m-1,2} & \cdots & \sigma_{m-1,m-1} & \sigma_{m-1,m} \\ \sigma_{m,1} & \sigma_{m,2} & \cdots & \sigma_{m,m-1} & \sigma_{m,m} \end{bmatrix} \quad (6)$$

Where m denotes the number of the random parameters, $\sigma_{k,j}$ represents the elements of the Cholesky matrix. The standard deviation of the correlated random parameters is based on the diagonal and off-diagonal elements of Eq. (6), which can be defined as follows:

$$\sigma_r=\sqrt{\sigma_{k,k}^2+\sigma_{k,k-1}^2+\sigma_{k,k-2}^2+\cdots+\sigma_{k,1}^2} \quad (7)$$

Where $\sigma_r$ represents the standard deviation of the random parameter, $\sigma_{k,k}$ is the Cholesky matrix's diagonal element($1\leq k\leq m$). The correlation coefficient between two random parameters was used to produce the correlations previously reported, which was given as:

$$\rho=\frac{Cov(X_1,X_2)}{\sigma_{X_1}\times\sigma_{X_2}} \quad (8)$$

Where $Cov(X_1,X_2)$ represents the covariance between $X_1$ and $X_2$ random parameters. $\sigma_{x1}$ and $\sigma_{x2}$ denote the standard deviations.

**4.2 Marginal effects**

To quantify the effects of categorical explanatory variable on independent variable outcome probabilities, the marginal effects are computed, as shown in Eq. (9)

$$M_{x_{in}}^{P_{in}}=P_{in}\left[\text{given } x_{in}=1\right]-P_{in}\left[\text{given } x_{in}=0\right] \quad (9)$$

Where $P_{in}$ [given $x_{in}$=1] is the probabilities of injury severity level i for observation *n* when the dummy variable equals one. Marginal effects are used to present the difference in the estimated probabilities when the categorical explanatory variables change from 0 to 1.

**4.3 Transferability test**

To examine if there are significant differences between front- and rear-car occupant group injury severity in rear-end crashes, the log-likelihood ratio (LR) test was employed between the full model and two sub-models(FV model and RV model). The null hypothesis considered there no significant difference between the two models. The test statistic $LR_{full}$ follows a $\chi2$ distribution (with degrees of freedom equal to the summation of the statistically significant parameters estimated in the two subset-data models minus the number of the statistically significant parameters estimated in the total-data model). The LR test can be expressed as Eq. (10):

$$LR_{full}=-2\left[LL(\beta_{full})-LL(\beta_{FV})-LL(\beta_{RV})\right] \quad (10)$$

Where $LL(\beta_{full})$ is the log-likelihood at the convergence of the overall model which can be obtained by using all the crash data; $LL(\beta_{FV})$ represents the log-likelihood at convergence of FV Model, and $LL(\beta_{RV})$ represents the log-likelihood at convergence of RV Model. The result of LR test is 175.4 and with 16 degrees of freedom, revealing that the null hypothesis can be rejected with>99.99% confidence. Therefore, front- and rear-car occupant injury severity should be modelled separately with a high level of statistical confidence.

**5 Results**

Initially, this paper aims to capture the association between random parameters by using two separate CRPLHM models for front-and rear-car occupant injury severity respectively. However, no significant association is found between the random parameters in the RV dataset. Hence, the random parameter logit model with heterogeneity in means (RPLHM) is used in the RV dataset. To verify the superiority of the model used, the Akaike Information Criterion (AIC), McFadden Pseudo R-squared, and Likelihood Ratio Tests are employed. The random parameter logit (RPL) model is used for comparison. The results indicate that McFadden Pseudo R-squared (and the AIC) for CRPLHM model in FV dataset is higher (was smaller) than the comparison model, which generally implies better statistical fit results. In RV dataset, the RPLHM model obtains better evaluation results. To determine whether the proposed the CRPOPHM model and RPLHM model generate a statistically significant improvement over other models, a series of chi-square distributed likelihood ratio tests are used.

As indicated in **Table 2**, the CRPOPHM model in FV dataset is statistically superior to all other models at a confidence level above 90%. The RPOPHM in RV dataset are statistically superior to all other models at a confidence level above 97%. In summary, considering the correlation among random parameters and heterogeneity in means can improve the statistical fit by a small but statistically significant amount (Se, C. et al., 2021). As such, the CRPLHM and RPLHM models are selected to analyze the injury severity of the front- and rear-car occupant groups respectively.

**5.1 Random parameter**

The model results are shown in **Table 2**. Only the independent variables that are statistically significant at the 90% significance level are presented (Ahmadi et al., 2018; Champahom et al., 2020). Five parameters (and three parameters) are found to be random and normally distributed in FV Model (in RV Model). Those random parameters (mean and standard deviation) have a different interpretation. For example, front-right position (the occupant at first row right position) is found to be statistically significant to serious injury with a mean of -1.474 and standard deviation of 3.286 in FV Model. The results indicate that driving position can reduce the likelihood of fatal injury in 67.36% of observations, while the probability of fatal injury increased in the remaining 32.64% of observations. This random parameter can capture the unobserved heterogeneity that may be caused by the headrest position, driver risk avoidance measures, crash speed, etc. In RV Model, young driver (front car driver is under 25 years) with a mean of -0.533 and standard deviation of 2.184 indicated that about 59.8% of observations model decrease the likelihood of fatal injury, while for 40.13%, the opposite is true. Their aggressive driving style and good physical fitness can be used to explain this heterogeneity. Moving to other parameters, the distribution effect of the random parameters across samples in both models is also shown in **Table 2**.

**5.2 Random parameter with heterogeneous means**

In FV Model, it can be witnessed that adverse weather is found increasing the means of lighted, and large truck. This suggests that the likelihood of fatal injury is decreased in adverse weather when cars rear-ended large trucks and rear-end collisions occurred at night on lighted roads. The young driver indicator (the rear vehicle driver is under 25 years) is found to increase the mean of lighted, which leads to an increasing likelihood of fatal injury. In addition, using safety belt is found to be negatively associated with the mean of the front-right position, implying that using safety belts in first row right position tends to reduce the possibility of fatal injury.

In RV Model, only one variable is significantly associated with the mean of the random parameters. The young driver indicator (the rear vehicle driver age is under 25 years) different from FV Model, is found to decrease the mean of large truck. This suggests that young drivers rear-ending large truck tend to decrease the probability of serious injury.

**5.3 Correlations among random parameters**

The CRPLHM model relaxes the assumption that the random parameters are uncorrelated. Only two random parameters are found to be significantly correlated in FV Model, i.e. decelerating, and large truck. This finding shows that the decelerating vehicles struck by large trucks tend to increase the possibility of fatal injury.

**5.4 Fixed parameters**

From occupant and crash perspectives, using seatbelts dramatically decreases the likelihood of fatal and serious injury in both models (by 23.36 %and 11.34% in FV Model; by 18.27% and 11.48% in RV Model), while the injury severity significantly increases in the driving position. It is unexpected that front-right position (the occupant at first row right position) as the random parameter is found to decrease the likelihood of fatal injury by 0.42% for 67.36% of observations in FV Model, but have positively associated with fatal injury in RV Model, which may be due to spatial instability and may once again suggest the necessity to segment the occupant by front and rear cars. In addition, decelerating (the rear-end crashes occurred due to the front vehicles decelerating) indicator is found to cause the worst consequences and significantly increase the likelihood of serious injury by 2.02% (fatal injury by 0.30% for 54.38% of observations) in FV Model, but not significant in RV Model. Slower (the rear-end crashes occurred due to the slower speed of the front vehicles) indicator is found to associate with a decreased likelihood of fatal injury in both models (by4.95% and 1.85%).

From environmental perspectives, only two variables are found to be statistically significant in the FV model. Crashes occurred at night with lights are highly associated with fatal injury with decreasing the probability by 3.24%. Asphalt roads are negatively correlated with fatal and serious injuries and reduce the likelihood by 4.24% and 4.96% respectively.

From vehicle perspectives, vehicle types are found to be played an important role in affecting occupant injury severity. Whether as the vehicle being rear-ended or being rear-ended, large trucks, SUVs, and pickups are found to be positively correlated with fatal injury. In particular, it has been observed that large trucks are the most dangerous vehicle type for car occupants involved in a rear-end crash, and the possibility of fatal injury is increased by 1.96% and 17.75% in FV and RV models, respectively. Vans are not significant in FV Model, but significantly increase the likelihood of fatal injury by 1.90% in RV Model. In addition, our study also found that one car rollover increased the probability of fatal injury to its occupants, while the other vehicle rollover reduced the probability to non-rollover car occupants. For instance, the indicator for the rear vehicle overturn is found to increases fatal injury probability by 1.72%, but front vehicle overturn is observed to decrease the likelihood of fatal injury by 2.65% in RV Model. This novelty finding once again demonstrates the necessity to divide the occupants into front-and rear-car groups.

From driver perspectives, it is interesting to note that older drivers cause a higher likelihood of fatal injury for their occupants, but also are less aggressive toward other vehicles in rear-end crashes. In FV Model, the likelihood of fatal injury increased by 2.65% when a passenger car driven by an older driver is rear-ended by another vehicle, but decreased by 2.76% is found when a passenger car is rear-ended by another vehicle driven by an older driver. The opposite results are found in RV Model. Young drivers are found to associate with a decreased likelihood of fatal injury in RV Model, but not significant in FV Model. Driver gender is a statistically significant factor in FV Model, but not significant in RV Model. The study found that occupants suffered the probability of fatal injury increased by 5.68% when the passenger car is rear-ended by a male driver. However, the passenger car driven by a male driver and rear-ended by another vehicle tends to reduce the probability of serious injury by 2.71%.

## Table 2 Estimation results and marginal effects

| Variables | FV Model | | | | | | RV Model | | | | | |
|---|---|---|---|---|---|---|---|---|---|---|---|---|
| | Fatal injury | | | Serious injury | | | Fatal injury | | | Serious injury | | |
| | *Coeff.* | *t-stat* | *Marginal* | *Coeff.* | *t-stat* | *Marginal* | *Coeff.* | *t-stat* | *Marginal* | *Coeff.* | *t-stat* | *Marginal* |
| Constant | .036 | .06 | | -3.440 | -3.80 | | -.959 | -1.87 | | .039 | .08 | |
| **Occupant and crash characteristics** | | | | | | | | | | | | |
| Safety belt (1if the occupant used the safety belt, 0 otherwise) | -2.295 | -6.19 | -.2336 | -1.242 | -3.27 | -.1134 | -2.505 | -7.59 | -.1827 | -1.117 | -4.36 | -.1148 |
| Driving position(1if the occupant at driver position, 0 otherwise) | .619 | 2.23 | .0424 | -1.893 | -1.95 | .0371 | 2.509 | 6.17 | .2047 | | | |
| Standard deviation | | | | 3.516 | 2.89 | | | | | | | |
| Front-right position (1 if the occupant at first row right position, 0 otherwise) | -1.474 | -1.86 | -.0042 | .720 | 2.64 | .0297 | 1.434 | 3.22 | .0265 | | | |
| Standard deviation | 3.286 | 2.58 | | | | | | | | | | |
| Stopped(1 if the rear-end crashes occurred due to the front vehicles stopping, 0 otherwise) | | | | .951 | 3.52 | .0387 | | | | | | |
| Slower(1 if the rear-end crashes occurred due to the slower speed of the front vehicles, 0 otherwise) | -.963 | -4.01 | -.0495 | | | | -.632 | -2.39 | -.0185 | | | |
| Standard deviation | | | | | | | .926 | 1.54 | | | | |
| Decelerating(1if the rear-end crashes occurred due to the front vehicle decelerating, 0 otherwise) | .265 | .59 | .0030 | 1.244 | 3.30 | .0202 | | | | | | |
| Standard deviation | 2.513 | 2.25 | | | | | | | | | | |
| **Environmental characteristics** | | | | | | | | | | | | |
| Lighted(1 if the rear-end crashes occurred at night with lights, 0 otherwise.) | -.509 | -1.13 | -.0324 | | | | | | | | | |
| Standard deviation | 2.362 | 2.70 | | | | | | | | | | |
| Asphalt (1if crash occurred at asphalt road, 0 otherwise.) | -.582 | -2.50 | -.0424 | -.769 | -3.10 | -.0496 | | | | | | |
| **Front vehicle and driver characteristics** | | | | | | | | | | | | |
| Older driver (1if the front vehicle driver age is above 60 years, 0 otherwise) | .883 | 3.39 | .0265 | | | | -0.547 | -2.06 | -.0118 | | | |
| Young driver (1if the front vehicle driver age is under 25 years,0 otherwise) | | | | | | | -.533 | -1.40 | -.0181 | | | |
| Standard deviation | | | | | | | 2.184 | 2.53 | | | | |
| Male (1if the front vehicle driver is male, 0 female) | | | | -.443 | -1.97 | -.0271 | | | | | | |
| Overturn (1if the front vehicle overturned, 0 otherwise) | .662 | 1.45 | .0071 | 1.229 | 2.73 | .0130 | -3.193 | -6.07 | -.0265 | -.803 | -3.46 | -.0224 |
| SUV (1if the front vehicle is a suv, 0 otherwise) | | | | | | | 1.277 | 3.88 | .0255 | | | |
| Pickup (1if the front vehicle is a pickup, 0 otherwise) | | | | | | | 2.016 | 5.32 | .0327 | | | |
| Van (1if the front vehicle is a van, 0 otherwise) | | | | | | | 1.491 | 2.79 | .0190 | | | |
| Large truck (1if the front vehicle is a large truck, 0 otherwise) | | | | | | | 4.379 | 8.64 | .1775 | .641 | .94 | .0581 |
| Standard deviation | | | | | | | | | | 2.169 | 2.41 | |
| Disable damage (1if the front vehicle is disable damage, 0 otherwise) | 2.396 | 5.57 | .2797 | 3.769 | 5.27 | .4062 | | | | | | |
| **Rear vehicle and driver characteristics** | | | | | | | | | | | | |
| Older driver (1if the rear vehicle driver age is above 60 years, otherwise) | -1.441 | -4.66 | -.0276 | | | | 1.742 | 5.11 | .0286 | | | |
| Young driver (1 if the rear vehicle driver age is under 25 years, otherwise) | | | | | | | -0.604 | -2.17 | -.0024 | | | |
| Male (1if the rear vehicle driver is male, 0 female) | .606 | 2.28 | .0568 | .989 | 3.21 | .0865 | | | | | | |
| Distracted driving (1if the rear driver is distracted driving, 0 otherwise) | .187 | .73 | .0051 | | | | | | | 1.121 | 2.50 | .1571 |
| Pickup (1if the rear vehicle is a pickup, 0 otherwise) | .637 | 2.41 | .0188 | | | | | | | | | |
| SUV (1if the rear vehicle is a suv, 0 otherwise) | .507 | 1.68 | .0125 | 0.662 | 2.10 | . 0147 | | | | | | |
| Large Truck (1if the rear vehicle is a large tuck, 0 otherwise) | 1.096 | 1.58 | .0196 | 0.605 | 1.55 | .0102 | | | | | | |
| Standard deviation | 4.011 | 2.17 | | | | | | | | | | |
| Overturn (1if the rear vehicle overturned, 0 otherwise) | -1.251 | -3.59 | -.0176 | | | | 1.520 | 3.70 | .0172 | | | |
| **Heterogeneity in means** | | | | | | | | | | | | |
| Front right positions: Safety belt (1 if the occupant used the safety belt, 0 otherwise.) | -.531 | -1.55 | | | | | | | | | | |
| Lighted: Young driver (1 if the rear vehicle driver is under 25 years, otherwise) | 1.734 | 1.87 | | | | | | | | | | |
| Lighted: Adverse weather (1 if weather condition were rain, snow, fog, 0 otherwise.) | 1.720 | 1.93 | | | | | | | | | | |
| Large Truck: Adverse weather (1 if weather condition were rain, snow, fog, 0 otherwise.) | 1.314 | 1.75 | | | | | | | | | | |

| | | | | | | | | |
|---|---|---|---|---|---|---|---|---|
| Large Truck: Young driver (1 if the rear vehicle driver age is under 25 years, otherwise) | | | | | | | -1.168 | -2.17 |
| Slower: Weekend(1 if the rear-end crash occurred on a weekend , 0 weekday) | | | | | .628 | 1.88 | | |
| | FV Model | | | | RV Model | | | |
| **Distributional effect of the random parameters (%)** | Fatal injury | | Serious injury | | Fatal injury | | Serious injury | |
| | Above zero | Below zero | Above zero | Below zero | Above zero | Below zero | Above zero | Below zero |
| Driving position | | | 70.54 | 29.46 | | | | |
| Front-right position | 32.64 | 67.36 | | | | | | |
| Lighted | 41.29 | 58.71 | | | | | | |
| Decelerating | 54.38 | 45.62 | | | | | | |
| Large truck | 60.64 | 39.36 | | | | | 61.79 | 38.21 |
| Young driver(front-car) | | | | | 59.8 | 40.2 | | |

Diagonal and off-diagonal matrix [t-stats], and correlation coefficients (in parenthesis) of random parameters (FV Model )

| | Decelerating |
|---|---|
| Large Truck | 9.939 [85.46](0.985) |

| **Model statistics** | FV Model | | | RV Model | |
|---|---|---|---|---|---|
| | CRPLHM | RPLHM | RPL | RPLHM | RPL |
| Number of observations | 1078 | 1078 | 1078 | 1002 | 1002 |
| Number of estimated parameters | 45 | 42 | 36 | 25 | 23 |
| Log-Likelihood at zero | -1184.3 | -1184.3 | -1184.3 | -1100.81 | -1100.81 |
| Likelihood at convergence | -956.14 | -959.36 | -969.45 | -733.22 | -737.21 |
| Pseudo R-squared | 0.1927 | 0.1899 | 0.1814 | 0.3339 | 0.3303 |
| AIC | 2002.3 | 2002.7 | 2010.9 | 1516.5 | 1520.4 |

| **Likelihood ration test** | FV dataset | | | RV dataset |
|---|---|---|---|---|
| | CRPLHM vs RPLHM | CRPLHM vs RPL | RPLHM vs RPL | RPLHM vs RPL |
| Degree of freedom | 3 | 9 | 6 | 2 |
| Resulting $\chi2$ | 6.44 | 26.62 | 20.18 | 7.98 |
| Level of confidence | 90% | 99% | 99% | 97% |
| Statistically superior model | CRPLHM | CRPLHM | RPLHM | RPLHM |

2

## 6. Discussion and conclusions

Model results show that using safety belts is associated with a lower likelihood of fatal injuries in both occupant groups. This finding is supported by Chen et al. (Chen, F. et al., 2019) that using safety belts is found to decrease the injury severity for both front and rear vehicle drivers in a rear-end crash. An extension of our study is that the use of seat belts is more efficient in protecting the front-car occupant group than the rear-car occupant group in a rear-end crash. This is understandable since when a rear-end crash occurs, the shock wave to the front-car occupant group is forward and the shock wave to the rear-car occupant group is rearward. Using safety belts can be used more effectively to mitigate the effects of forward shock waves. To promote the use of seat belts, a straight-forward approach is to upgrade secondary seat belt laws to primary seat belt laws. In the United States, there are still 15 states that enforce secondary seat belt laws. Boakyea and Nambisan(Boakye & Nambisan, 2020), found that upgrading seatbelt laws from no/secondary laws to primary laws covering all vehicle occupants will have a positive impact on seatbelt use. Another interesting conclusion is that seating position is found to have a strong correlation with the occupant's injury severity in both models. In particular, driving position is found to be more dangerous than the other position. This finding is contrary to Bogue et al.(Bogue et al., 2017), who analyzed the injury severity of occupants in multivehicle crashes and found the front-right seat position is more dangerous than the driving position. One possible explanation is that the buffer space in the car driving position is smallest thus increasing the probability of fatal injuries. In addition, the driving and front-right positions are more dangerous in RV Model than in FV Model. One reason for this is that the seat mitigates the impact of the rearward shock wave but not the forward shock waves. Therefore, it is recommended that all occupants should be required to use seat belts.

Model findings reveal that the most dangerous rear-end crash trigger is the sudden deceleration of the car, especially when followed by a large truck. High speed differentials and short evasion times can be used to explain this phenomenon. On the one hand, we suggest strengthening technical training for car drivers, for example, by providing traffic safety education for illegal drivers, by improving their safety awareness, and by increasing the difficulty of obtaining a driver's license for the second time. On the other hand, Advanced Driver Assistance Systems (ADAS) can be used to help drivers improve their driving habits.

Asphalt road has a negative effect on occupant injury severity in FV Model. The result is supported by Yuan et al.(Yuan, Y. et al., 2021). The finding suggests that upgrading concrete or others pavement to asphalt can effectively reduce rear-end crash severity. Rear-end crashes occurring at nighttime with light conditions reduce the likelihood of fatal injury. This finding is supported by Obeidat and Rys(Mohammed Said Obeidat) who investigated the benefits of intersection lighting in both rural and urban areas, and found a 3.61% and 6.54% reduction in nighttime crash frequency compared to dark intersections. One possible reason for this could be that drivers can quickly identify rear-end risks in brighter light conditions. It may be helpful to increase the frequency of inspection and maintenance of street lights, while further improving the coverage of the street lights.

Large truck is the most dangerous vehicle type vehicle for car occupants involved in a rear-end crash. This finding may be supported by the fact that the longer braking distances and shorter reaction times of larger trucks, which always lead to a higher speed before the crashes. Therefore, lower speed limits for large truck should be implemented. This study also reports that large trucks driving under adverse weather could reduce the probability of the fatal injury in RV Model. This phenomenon can be explained by the finding of Wang and Prato (2019) that large truck drivers driving more carefully at lower speeds in adverse weather. Moreover, a large truck rear-ending a passenger car is more dangerous than a large truck rear-ended by a passenger car. This finding implies that large trucks should maintain a longer following distance from passenger cars. This result is partly agreed with Yuan et al (2022), who have found large trucks striking other vehicles can increase crash severity, but being struck can lead to contrary consequence at unsignalized intersections.

Many previous studies have reported that crashes involving overturns tend to cause severe injuries(Peng, Y. et al., 2018; Wang, Y. & Prato, 2019). Interestingly, our study fills this gap by finding that one car rollover increased the probability of fatal injury to its occupants, while the other vehicle rollover reduced the probability to non-rollover car occupants. This finding suggests that when faced with the crash risk, drivers should avoid overreacting to reduce the likelihood of overturning. It is essential to improve the driver's emergency response ability and psychological quality, such as through psycho-education, awareness campaigns, and mandatory workshops.

Older drivers cause a higher likelihood of fatal injury for their occupants, but also are less aggressive toward other vehicles in rear-end crashes. This result can be used to explain the controversial conclusion mentioned about older drivers in Section 2. A likely explanation for this association between "older drivers" and the death of their passengers is the frailty of their passengers: older drivers tend to drive older passengers (or sometimes young children). Cognitive limitations also increase the odds of crash involvement for older drivers, but their extensive driving experience may help them reduce the crash severity(Islam & Burton, 2019). One potential application is that display signs in vehicles driven by older drivers to remind other vehicles maintaining a greater following distance. It is expected that young drivers for their shorter reaction time have negative effect on occupant injury severity in RV Model. This result is supported by Chen et al. (Chen, F. et al., 2019). Nevertheless, young drivers are found to increase the likelihood of fatal injury at nighttime on lighted roads. This finding could be explained by the fact that young drivers driving at high speeds are more likely to relax their vigilance on lighted roads. Driver gender is a statistically significant factor in FV Model, but not in RV Model. Dabbour et al. (Dabbour et al., 2020)analyzed the contributing factors affecting driver injury severity in rear-end collisions and found that the injury severity of struck drivers decreases if the striking driver is female, but increases if the struck driver is female. Interestingly, our study further expanded boundaries of knowledge by finding that struck occupant injury severity tend to increase when the striking driver is male, but decrease when the struck driver is male in rear-end crashes. This finding suggests that male drivers should reduce speed when following a car. It is expected that the disabling damage of vehicle are more likely to sustain more severe injury in both models, in which is consistent with the study of Chen et al. (Chen, C., Zhang, Yang, et al., 2016).

The main contribution of this paper is that investigate differences between front-and rear-car occupant group injury severity involved in fatal rear-end crashes. To determine if passenger car occupant groups should be considered separately, the log-likelihood ratio (LR) test is conducted; the results suggested that front- and rear-car occupants should be modelled separately with a high level of statistical confidence. Furthermore, the CRPLHM models are developed to analyze different occupant group injury severity respectively. The estimated model results indicate that there are no significant correlations among the random parameters in FV Model, and thereby a mixed logit model with heterogeneity in means (RPLHM) is used to analyze the front-car occupant group injury severity. Additionally, the RPL, RPLHM, and CRPLHM models are estimated and the fitting performances are compared. Various risk factors associated with two occupant group injury severity are identified, and these results are helpful to guide traffic managers to implement corresponding countermeasures to improve passenger car safety.

Although some results are concluded, there are still limitations in this study. Firstly, as some of the sample variables are missing, only a limited sample is selected. Secondly, some factors that may affect crash injury severity such as vehicle speed, traffic volume, etc, are not considered in this paper due to the limitation of the data. For future studies, there are also several promising extensions of this work, e.g., addressing temporal correlation in injury severity, examining different collision types, or examining different vehicle types. The extraction of current traffic accident samples can be influenced to some extent by the personnel involved in the process. ChatGPT has started to generate utility in the fields of medicine and transportation (Huang et al.,2023; Zheng et al.,2023). In the future, it can be considered to utilize ChatGPT for conducting relevant research in this area.

**Acknowledgments**

This research was funded by the National Natural Science Foundation of China (No. 71871059), Postgraduate Research &Practice innovation Program of Jiangsu Province (No.KYCX22_0270)